\documentclass[submission,copyright,creativecommons]{eptcs}
\providecommand{\event}{FESCA2014} 
\usepackage{breakurl}             
\usepackage{graphicx}
\usepackage[table]{xcolor}

\title{Transformation of UML Behavioral Diagrams to Support Software Model Checking}
\author{Luciana Brasil Rebelo dos Santos \and Valdivino Alexandre de Santiago J\'unior \and Nandamudi Lankalapalli Vijaykumar
\institute{Laborat\'orio Associado de Computa\c{c}\~ao e Matem\'atica Aplicada - LAC\\
Instituto Nacional de Pesquisas Espaciais - INPE \\
S\~ao Jos\'e dos Campos - SP, Brazil}
\email{luciana.santos@lac.inpe.br \quad valdivino.santiago@inpe.br \quad vijay.nl@inpe.br}
}
\def\titlerunning{UML Behavioral Diagrams to Transition System}
\def\authorrunning{L.B.R. Santos, V. A. Santiago Jr., N.L. Vijaykumar}
\begin{document}
\maketitle

\begin{abstract}
Unified Modeling Language (UML) is currently accepted as the standard for modeling (object-oriented) software, and its use is increasing in the aerospace industry. Verification and Validation of complex software developed according to UML is not trivial due to complexity of the software itself, and the several different UML models/diagrams that can be used to model behavior and structure of the software. This paper presents an approach to transform up to three different UML behavioral diagrams (sequence, behavioral state machines, and activity) into a single Transition System to support Model Checking of software developed in accordance with UML. In our approach, properties are formalized based on use case descriptions. The transformation is done for the NuSMV model checker, but we see the possibility in using other model checkers, such as SPIN. The main contribution of our work is the transformation of a non-formal language (UML) to a formal language (language of the NuSMV model checker) towards a greater adoption in practice of formal methods in software development.
\end{abstract}

\section{Introduction}

Verification and Validation (V\&V) play a key role of getting quality and has been gaining much importance in the academia and private sector. 
Formal methods offer a large potential to obtain an early integration of verification in the design process, and to provide more effective verification techniques \cite{baier/08}. 
 However, formal methods require mathematical background and their use is restricted, as users prefer the simplicity of other notations, rather than more formal means. Thus, adoption of formal methods will be easier when they can be applied within standard development process and when they are based on standard notation \cite{schafer2001model}.

Unified Modeling Language - (UML) \cite{umlref} is currently accepted as the standard for modeling (object-oriented) software. It presents diagrams that represent the static structure of a system, and also defines diagrams to model the dynamic behavior of systems. 
In particular, dynamic aspects of system behavior can be specified by interactions (i.e. sequence diagrams); UML behavioral state machines (variant of Harel's Statecharts) and activity diagrams give a view of the system that is associated with instances of classes. These types of diagrams represent complementary views of the system, but, at the same time, hide redundant descriptions of the same aspects of the system. This gives the opportunity for V\&V techniques to ensure the consistency of these descriptions and the system requirements.

This paper presents an approach related to Model-Driven Development and Formal Verification. Our approach transforms up to three different UML behavioral diagrams (sequence, activity, behavioral state machines) into a single Transition System (TS) to support Model Checking of software developed in accordance with UML. We consider properties generated from use case descriptions, which represent the requirements, and the TS translated from the behavioral diagrams. Our verification essentially consists of sequence of scenarios to be checked.
Besides, the single TS has a unified view of different perspectives of behavioral modeling of the system, obtained by using various UML diagrams. The main contribution of our work is the transformation of a non-formal language (UML) to a formal language (language of the NuSMV model checker) towards a greater adoption in practice of formal methods in software development.

The rest of the paper is organized as follows. In the next section we explain our approach to apply Formal Verification. This section also shows the details for constructing the TS, and for generating the model checker notation. In Section 3 we apply our approach to a case study and shows preliminary results. Related work is discussed in Section 4. Finally, Section 5 concludes the paper.

\section{Transforming UML behavioral diagrams into Transition Systems (TS)}

A prerequisite for model checking is a property to be checked and a model of the system under consideration. In Figure~\ref{solimva3}, we show our approach aiming at Model Checking of software developed in accordance with UML. In the following, we explain the main activities that the workflow performs: 

\begin{figure}[!h]
\centering
\includegraphics[width=0.40\textwidth]{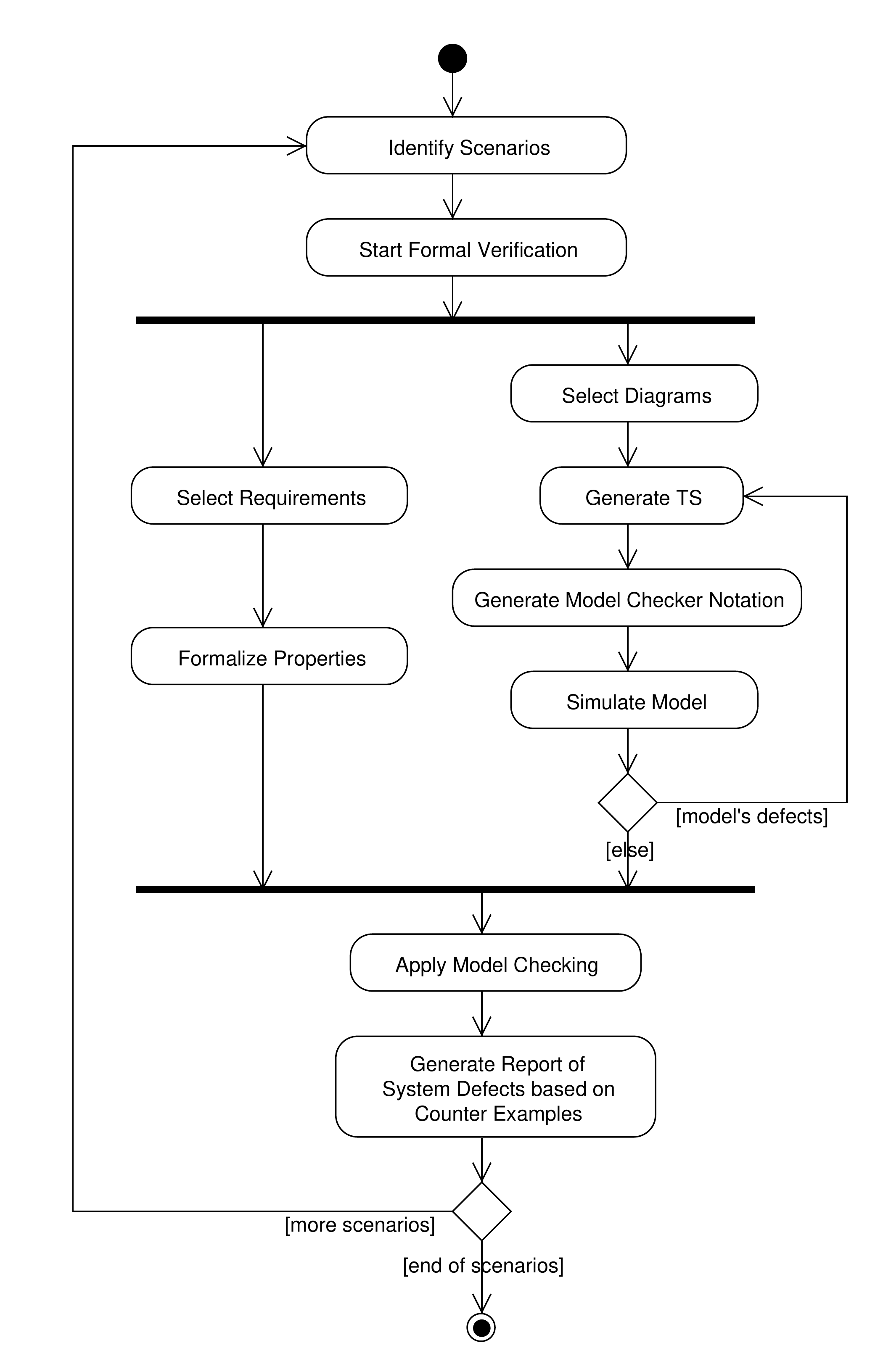}
\caption{Workflow of the proposed approach}
\label{solimva3}
\end{figure}

(i) \textbf{identify scenarios} by looking at use case models. A use case can be viewed as a scenario. Each scenario is a set of related subscenarios tied together by a common goal. The mainline sequence ('main success scenario' \cite{cockburn}) and each of the variations ('extensions and sub-variations') are the scenarios identified by our approach; (ii) \textbf{formalize properties}. For each selected scenario, we extract requirements from the textual description of use cases. After that, we formalize properties by means of specification patterns \cite{dwyer/99}; (iii) \textbf{generate TS}. Based on the available UML behavioral diagrams, we generate a single TS (finite-state model). This is the main activity to achieve the objective of our work. As will be explained later, our approach does not demand that all three UML behavioral diagrams (sequence, activity, behavioral state machines) exist: it is enough to have the sequence diagram and one of the other two to generate the TS; (iv) \textbf{generate model checker notation}. Then, we translate the created TS to a model checker. Our first idea is to use the NuSMV model checker \cite{NuSMV}; (v) Finally, we \textbf{apply Model Checking} to realize about defects in the behavioral description of the system represented by the UML diagrams.
We repeat activities from (ii) to (v) for each selected scenario. Also note that activities (ii) and (iii)/(iv) may be accomplished in parallel. 

In the next subsection we will detail the main activity of our approach, i.e. a solution to generate a single TS based on UML behavioral diagrams. We will focus on activity (iii) described above. It works in two phases: in the first phase, the single TS from each one of the diagrams is generated separately. Due to space constraints, we refrain to show the construction of individual TSs from each one of the diagrams. We focus on the second phase, which shows how to obtain the definitive/unified TS from the combination of the individual TSs. Finally, in Section \ref{nusmv} we present directives for the generation of the NuSMV notation (activity (iv)).

\subsection{The Unified Transition System}
In this subsection, we show how to generate the final/unified TS. Some definitions and notations used in our approach are given in the sequence.
For brevity, we denote Sequence diagram as SD, Behavioral State Machine diagram as SMD, and Activity diagram as AD. 
As previously mentioned, a scenario is identified by looking at use case models. We take for granted that every use case has at least one SD describing it.
We also assume that a scenario must have at least two diagrams describing it: one is mandatory, the SD , and the other one can be any of the other two diagrams. 

A state in the TS is identified by a tuple $\ll(Message,State,Activity),$ $g0,g1,...,gn\gg$ where \textit{Message} is from SD; \textit{State} is from SMD; and \textit{Activity} from AD; \textit{g0,g1,...,gn} represent all the existing guards in the diagrams along with their respective values. At the beginning, all guard values are assigned to '\textit{dc}', which means 'do not care', because at the beginning we don't know the guard values. If there are parallel messages, states, or activities, an '\textit{and}' is inserted in the tuple, such as $\ll(Msg1andMsg2,State1andState2,Activity1andActivity2),...\gg$, which means that \textit{Msg1} and \textit{Msg2} are sent in parallel \textit{State1} and \textit{State2} are parallel states, as well as \textit{Activity1} and \textit{Activity2}. When a diagram is missing, its part in its corresponding state in the TS is substituted by '\textit{-}'. It also happens when there is some behavior that is not modeled in one of the diagrams, that is, when the diagrams are incoherent or when there is a more detailed description in one of them. In these situations, states are described as $\ll (Message,State,-), g0,g1,...,gn \gg$ or $\ll (Message,-, State),$ $g0,g1,...,gn \gg$.

The main challenge to combine the three diagrams is to define the correspondence among messages, events, and activities. For instance, an event for a transition of the SMD can be a message of the SD, while the message can be a call to an activity modeled by an AD. Nevertheless, finding where a specific message of the SD is modeled in the SMD or in the AD is very hard. Instead, we build the final TS using the individual TSs. We look over the TSs starting from their initial states. The correspondence among states of each TS is assigned by using the flow of transitions. At each iteration, we parse the three TSs and construct the new states by taking the next possible transitions in each TS. 

The flow of transitions, alone, is not enough to construct the final TS, as each TS contains multiple paths. In addition, we use guard values to help find the correspondence among states.
As we have already stated, the SD is mandatory and conducts the diagrams merging. The basic elements that delineate the possibility of paths in the SDs are guards in the combined fragments. Thus, we adopt the guard values along with the flow of transitions to construct the final TS states. 
If we think of an algorithm, at each iteration, the algorithm seeks the next possible transitions in each TS and its guard values.
To find out states that have the same guard values, we have created the \textit{gvs} (guard value structure) which represents the values of each one of all available guards on each state. 
Every time, for each possible next transitions of the TSs created from the SD, SMD, and AD, their \textit{gvs} are verified and a match of its values is created. Thereafter, new states are generated.

\begin{figure}[h]
\centering
\includegraphics[width=0.67\textwidth]{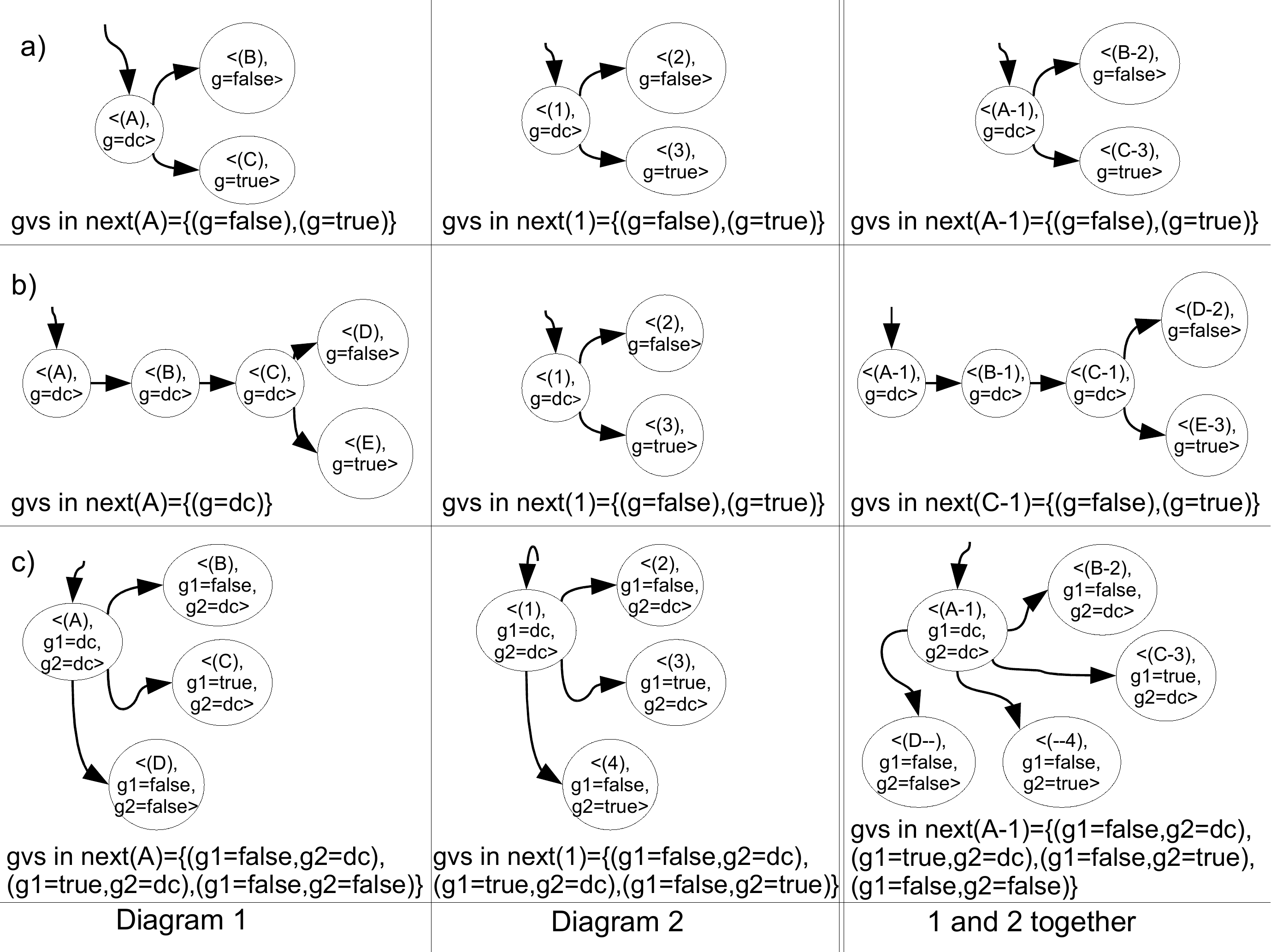}
\caption{Possible situations to generate the unified TS and its respective gvs}
\label{gvs}
\end{figure}

Figure \ref{gvs} shows the main possible situations that may occur to generate a state in the final TS. The diagram in column 3 is the combination of the diagrams in columns 1 and 2. Situation a) is straightforward to see. Situation b) shows how the algorithm handles cases when in at least one of the next possible states, the value of its \textit{gvs} is the same of the current \textit{gvs} value. This occurs because the diagrams do not model the same behavior during their evolution. An SMD may contain more details than an SD, and so on. In this case, the TSs have different lengths. Situation c) shows an example of inconsistent states, due to inconsistent diagrams. As previously discussed, this may happen when one of the diagrams models a behavior that is not modeled in the other diagram and vice versa. New transitions are created, as many as necessary to cover all possibilities of \textit{gvs} values.

There is still a situation we did not mention in Figure \ref{gvs}: TSs with no guards. Here we have two possibilities. The first one is the case of all diagrams have no guards. We assume that when a diagram has no guards, this means that there is no multiple paths. For example, a sequence diagram with no combined fragments, or with only parallel combined fragment. All other combined fragments require guards. Or an activity diagram with no decision node. 
In the absence of multiple paths, it is quite easy to construct the final TS: we only need follow the flow of transitions. The second one is the case of a diagram has guards and the other one does not. At the beginning, the algorithm seeks all guards in all diagrams. Suppose TS1 has guards \textit{g1} and \textit{g2} and TS2 has no guards. The \textit{gvs} will be composed by $\{g1,g2\}$ and it is attributed to TS2. As originally TS2 did not contain \textit{g1,g2}, their values are assigned as "\textit{dc}" and will never change. Thenceforth the situations are covered in Figure \ref{gvs}.

\subsection{Generation of model checker notation} \label{nusmv}

After its creation, the unified TS can be used to systematically generate its corresponding encoding into the model checker input language by constructing declarative divisions \cite{debbabi/10}. It is important to emphasize that once a formal unified TS was generated from UML behavioral diagrams, we see the possibility of transforming it into several different languages of available model checkers such as SPIN \cite{holzmann2004spin}, UPPAAL \cite{behrmann2004tutorial}, and NuSMV \cite{NuSMV}. 

In our current approach, we chose the NuSMV model checker because it is open source, it has a widespread use in academia, and it accepts properties formalized not only in Computation Tree Logic (CTL) but also in Linear Temporal Logic (LTL) \cite{baier/08}, two logics that are well known and have mapping defined in the specification patterns \cite{dwyer/99}. 

Considering the NuSMV model checker, declaration of variables is relatively easy. One variable we call $State$ which is related to the element  $\ll$\textit{(Message-State-Activity)}$\gg$ of the tuple that identifies a certain state of the TS. In addition, there will be as many variables as the guards identified within the UML behavioral diagrams, i.e. $g0$ is transformed into a variable $vg0$, $g1$ into a variable $vg1$, and so on. All these variables derived from the guards will be enumerated with the following values: $\{dc,false,true\}$. The $dc$ value is important because in many occasions and especially at the beginning of the behavioral system modeling, the values of certain guards or do not care or do not make sense be true or false. Another remark is that we use only state variables of NuSMV. 

The initial value of the $State$ variable of the final TS in NuSMV is always \textit{Start-Start-Start}. Note that this is not the "full" state characterization of the initial state of the NuSMV model because we depend on the values of the variables representing the guards to know this. To generate the model in NuSMV we simply follow the transitions of the unified TS, making appropriate assignments to the $State$ variable and for each variable derived from the guards.

\section{Preliminary Results}

\begin{figure}[!h]
\centering
\includegraphics[width=0.9\textwidth]{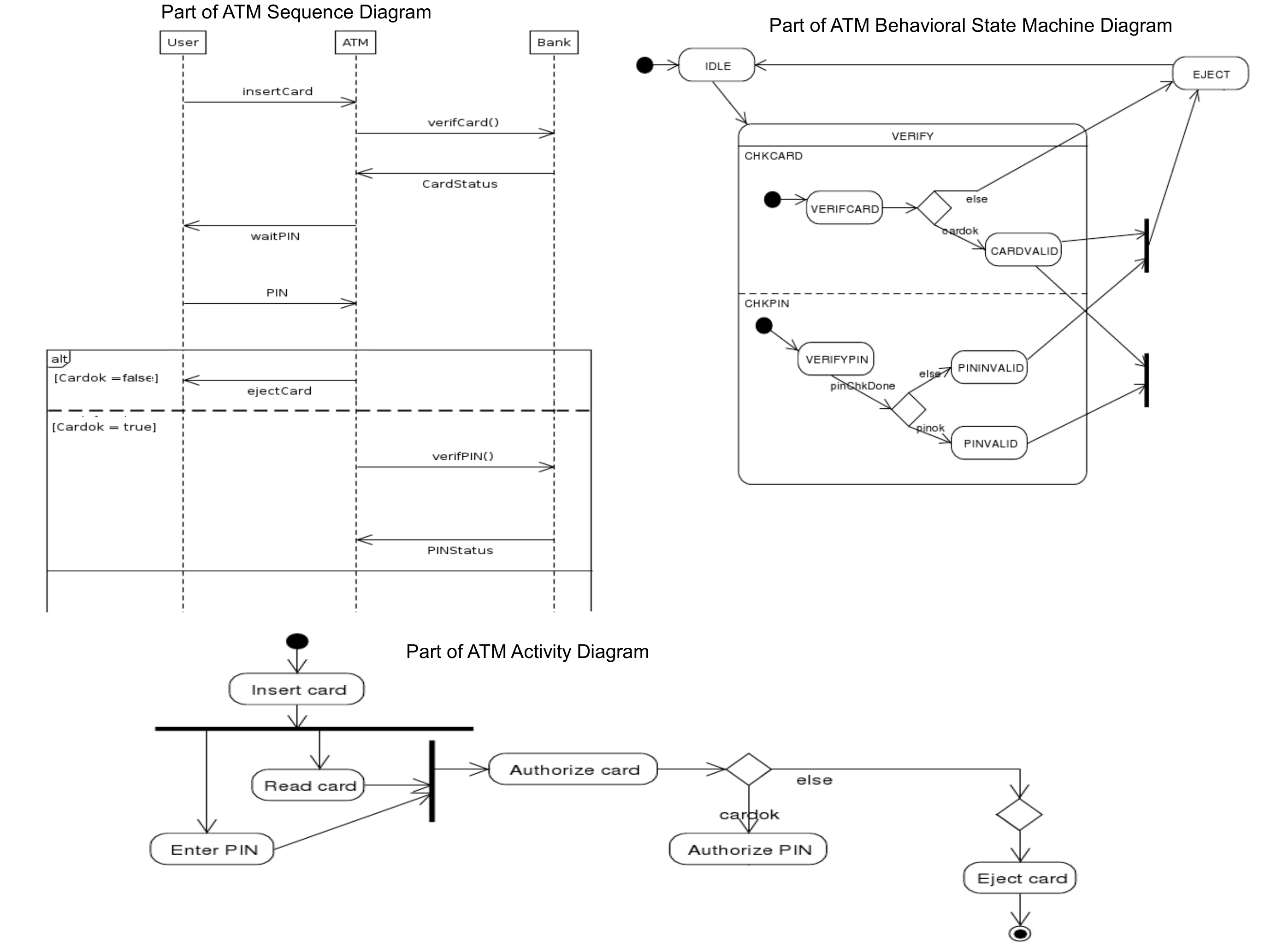}
\caption{Initial part of UML diagrams for ATM case study }
\label{diagrams}
\end{figure}

This section presents a case study to illustrate the feasibility of our approach.
We consider more thoroughly the ATM case study, where the ATM interacts with a customer via a specific interface and communicates with the bank over an appropriate communication link. 
The initial part of the sequence, activity and behavioral state machine diagrams for ATM are illustrated in Figure \ref{diagrams}. 

\begin{figure}[!h]
\centering
\includegraphics[width=0.66\textwidth]{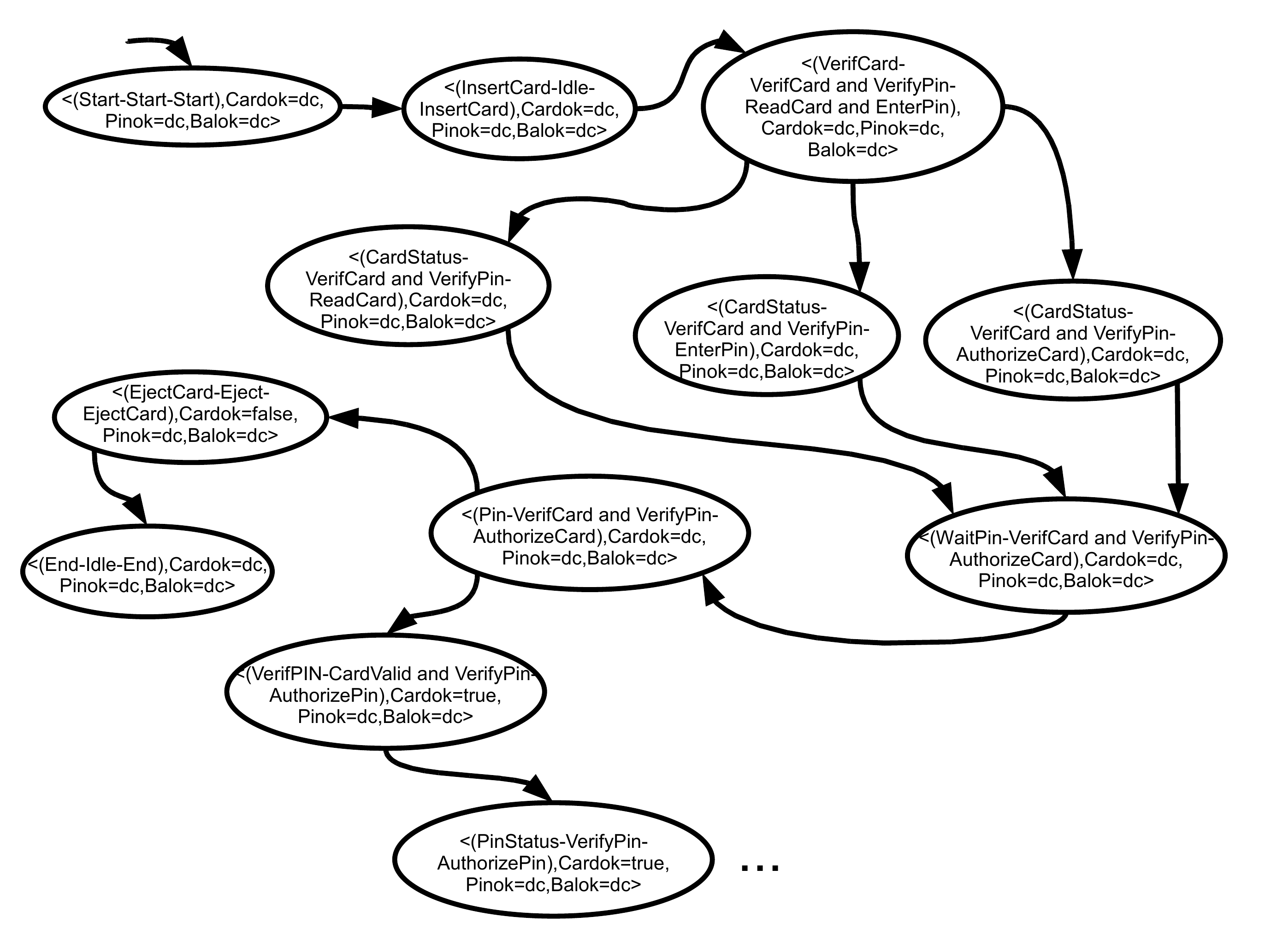}
\caption{Part of the final TS obtained from the three diagrams}
\label{ts}
\end{figure}

In accordance with our approach, we must \textbf{identify the scenarios}, observing the use cases. Then, we \textbf{start Formal Verification}. For this, we \textbf{generate TS} and \textbf{select requirements}.
Applying the suggested approach manually on the three diagrams, we achieve the TS shown in Figure \ref{ts}. Actually, Figure \ref{ts} exhibits only part of the unified TS (only the part related with Figure \ref{diagrams}). In total, we have 756 states, 86 of which are reachable states, when running NuSMV model checker. Each state is characterized by the values of the variables. We have identified four main variables that characterize the TS:
(i) \textit{State = \{Start-Start-Start,...,End-Idle-End\}}; (ii) \textit{CardOk = \{dc,false,true\}}. \textit{CardOk} represents the card validation; (iii) \textit{PinOk = \{dc,false,true\}}. \textit{PinOk} represents the PIN validation; and (iv) \textit{BalOk = \{dc,false,true\}}. \textit{BalOk} represents the available amount of the customer account. \textit{CardOk, PinOK} and \textit{BalOk} comes from the guards identified within the UML diagrams.

Continuing the approach, considering the use case \textit{Perform Transaction}, we can extract two relevant user-defined properties related to requirements.
To proceed with the Formal Verification, it is necessary to \textbf{formalize the properties} to be checked. We chose Computation Tree Logic (CTL) \cite{baier/08} to formalize the properties. Note that the properties could be formalized using LTL as well, considering that NuSMV support such logic.

 \textbf{Requirement 1}: \textit{the customer can perform transactions only if he/she has a valid card and a valid personal identification number (PIN). Otherwise, he/ she can not perform any kind of transaction}. This property can be formalized using the \textbf{Absence Pattern and Scope After Q} proposed by \cite{dwyer/99}, in CTL, as follows\footnote{We used the NUSMV's syntax to write the property in CTL.}:\\
\footnotesize
$AG (( CardOK = false \quad | \quad PinOK = false) \rightarrow AG (!(State = WaitAccount-CardValidandPinValid-InitiateTransaction)))$\\
\\
\normalsize
\textit{CardOk} and \textit{PinOk} refer to the validation of the customer credentials. The first state on the final TS which can be reached only when \textit{CardOk} and \textit{PinOk} are \textit{true} is \textit{WaitAccount-CardValidandPinValid-InitiateTransaction}, that is, when the custumer has card and PIN valids. If we observe the individual TSs, we can see that state \textit{WaitAccount} for the SD is the first reachable state when both \textit{CardOk} and \textit{PinOk} are \textit{true}, for SMD \textit{CardValidandPinValid}, and also \textit{InitiateTransaction} for AD.\\
\textcolor{white}{aaaa}\textbf{Requirement 2}: \textit{whenever the specified amount exceeds the level of available funds, it should be possible for the user to request a new cash advance operation if the user wishes to correct the amount}. The property can be formalized using the \textbf{Existence Pattern and Scope After Q} proposed by \cite{dwyer/99}, in CTL, as follows:\\
\footnotesize
$A[!(State = InsuffFunds-Modify-ShowBalance)  W ((State = 
InsuffFunds-Modify-ShowBalance) \& AF(State =\\ CashAdvance-Chkbal-CheckBalance))]) $\\
\\
\normalsize
The state that allows the customer to perform a cash operation is \textit{CashAdvance-Chkbal-CheckBalance} and the state that indicates that the available funds are insufficient is \textit{InsuffFunds-Modify-ShowBalance}. To formalize the properties, it is necessary to see the final TS and its variables.

The next task is to \textbf{generate the model checker model} based on the previous generated final TS. The guidelines to do this were presented in Section \ref{nusmv}. The model checker code will be automatically generated when the approach is implemented.
A small part of the NuSMV source code is shown below for this case study.

\scriptsize

\textcolor{white}{aa}\\
\texttt{MODULE main} \\
\texttt{VAR} \\
\textcolor{white}{aa} \texttt{State: \{Start-Start-Start,InsertCard-Idle-InsertCard,...,End-Idle-End\}};\\
\textcolor{white}{aa} \texttt{CardOk: \{dc,false,true\}};\\
\textcolor{white}{aa} \texttt{PinOk: \{dc,false,true\}};\\
\textcolor{white}{aa} \texttt{BalOk: \{dc,false,true\}};\\
\texttt{ASSIGN} \\
\textcolor{white}{aa} \texttt{init(State):= Start-Start-Start}; \\
\textcolor{white}{aa} \texttt{next(State):= case} \\
\textcolor{white}{aaaa} \texttt{State=Start-Start-Start \& CardOk=dc \& PinOk=dc \& BalOk=dc : InsertCard-Idle-InsertCard};\\
\textcolor{white}{aaaa} \texttt{State=InsertCard-Idle-InsertCard \& CardOk=dc \& PinOk=dc \& BalOk=dc : }\\
\textcolor{white}{aaaaaaaaaaa} ...\\
\textcolor{white}{aa} \texttt{esac;}\\

\normalsize

After applying Model Checking, the results indicate that the first propriety is satisfied. However, the second property is violated, generating a counterexample. When analyzing the counterexample, one can note that it is not possible to reach state \textit{CashAdvance-ChkBal-CheckBalance} (where is possible to request a new cash advance operation) from state \textit{InsuffFunds-Modify-ShowBalance} (where available funds are insufficient).

%
%

In this section, we have shown the feasibility of our approach, by means of a traditional case study performed manually, to generate the unified TS from the various UML behavioral diagrams, and we introduced policies to transform the TS into the NuSMV input language. Besides, after running the model checker, we have found that the behavioral modeling described in the UML diagrams does not comply with all the specified requirements.

\section{Related Work} \label{relw}

This section presents some of the research literature related to this paper, showing (not exhaustive) approaches that use Formal Verification and UML.

Knapp \cite{schafer2001model} used two complementary UML notations for the specification of dynamic system behavior - state machines and collaborations - to automatically verify whether the interactions expressed by a collaboration can indeed be accomplished by a set of state machines. The approach applied consistency checking between diagrams.

Mikk \cite{mikk1998implementing} translated Statecharts into PROMELA, the input language of SPIN verification system. 
Lam \cite{lam2007formalism} formally analyzed activity diagrams using NuSMV model checker. The objective was determining the correctness of activity diagrams. Eshuis \cite{eshuis2006symbolic} presented two translations from activity diagrams to NuSMV. The aim was to assess the activity diagrams from the point of view of requirements and also from the point of view of implementation, which represents the current system behavior. 
In \cite{anderson1996model}, Anderson translates the specification of TCAS (Traffic Alert and Collision Avoidance System), which is specified in RSML (Requirements State Machine Language) into the input language of NuSMV. The objective was to investigate if Model Checking could be used in large software specifications. Dubrovin \cite{dubrovin2008symbolic} implemented a tool that translates UML hierarchical state machine models to the input language of NuSMV too. Uchitel \cite{uchitel2001workbench} proposes translation of scenarios, specified as Message Sequence Charts (MSCs), into a specification in the form of Finite Sequential Processes. This can then be fed to the Labelled Transition System Analyser model checker to support system requirements validation. All these previous studies deal with a single UML or UML-like diagram to perform Formal Verification. Rather, our approach allows to work with up to three UML behavioral diagrams. In addition, it is not clear if in the previous studies the authors used specification patterns to formalize the properties. Specification patterns provide clear guidelines to such formalization. We are proposing not only a tool to translate UML diagrams into a unified TS to support Model Checking but also a full approach to detect defects in the design of software developed in accordance with UML.

Baresi \cite{baresi/12} developed a tool to carry out Formal Verification of UML-based models, mainly interested in the timing aspects of systems. It is composed by: static part (class diagrams); dynamic aspects and behavior are rendered through: (a) state diagrams and activity; (b) sequence diagrams; and (c) interaction overview diagrams, used to relate different sequence diagrams;
Clocks (and time diagrams) are used to add the time dimension to systems. All these diagrams seem to be required to construct the approach.

Cortellessa \cite{cortellessa2002prima} proposes a methodology Performance Incremental Validation in UML (PRIMA-UML) aimed at generating a queueing network based performance model from UML diagrams that are usually available early in the software lifecycle (use case, sequence, and deployment).
Bernardi \cite{merseguer2002compositional} translated sequence and statechart diagrams into Generalized Stochastic Petri Nets.
Both works aimed analyze performance aspects of systems. 
Our approach is more related to functional aspects of the software product.

Calinescu \cite{johnson2013incremental} used a probabilistic model checker (PRISM) to verify critical systems, after changes. 
Verifying these software systems only at design time is insufficient, they have to be reverified after each change. They did not work with UML diagrams, but with components and deterministic finite automata.

The main motivation of our approach is the practical use of formal methods in software development, through automation. The idea is that the user feeds the tool with UML behavioral diagrams and it shows the defects. This can be done throughout the lifecycle, even before software coding. \cite{cortellessa2002prima} suggested an interesting approach to encompass performance validation task as an integrated activity within the development process. We aim at detecting design defects within the solution, but considering functional requirements of the software product.
Besides, our approach suggests to get a single vision of the system captured from three different diagrams (sequence, activity, and behavioral state machines). These diagrams represent complementary views of system behavior and are often used in different phases of software specification and design, allowing thus a wider system range to be verified. Most of the works we mentioned deal with a specific type of UML diagram. \cite{baresi/12} seems to require an assorted number of diagrams (structural, behavioral and time diagrams), which are not always available on the documentation. Adversely, our proposal requires two diagrams as mandatory (one is sequence and another activity or behavioral state machine), which provides a higher chance of being used in real applications.

\section{Conclusions}

In this paper we presented an approach that, ultimately, is another initiative in order to facilitate and thus increase the use of formal methods in software real projects. We draw on two facts to the development of this work. First, UML is a language widely used in various application domains, including the aerospace one. Second, Formal Verification and formal methods in general, despite all the benefits already presented by academic community, have not seen widespread adoption as a routine part of systems development practice \cite{woodcock/09}.

We presented an approach that transforms up to three UML behavioral diagrams (sequence, activity, behavioral state machines) into a single Transition System to support Model Checking. Our proposal requires only 1 diagram as mandatory (sequence) and another (activity or behavioral state machine) in order to create the TS. If there are the 3 diagrams, our approach can also generate the unified model (TS). Thus, we believe that it has a great potential to be used in real systems development because not always a significant variety of UML diagrams is within the artifacts (requirements specification, software design document) created to develop certain type of software.

Future directions follow. We are currently developing a tool to support this transformation. We have already developed the transformation from SD into a single corresponding TS, and we are working at the AD transformation.
When SMD transformation is finished we will begin the construction of the final part, the unified TS generation. Another direction is to automatically identify in the UML diagrams a problem (inconsistency between diagrams, incorrect behavior) when a counterexample is detected by running NuSMV. We will dedicate efforts in transforming the unified TS to other model checkers such as SPIN and UPPAAL. More complex case studies in several domains such as space software \cite{santiago/12} and general purpose software are in our targets.\\

\footnotesize
ACKNOWLEDGEMENT\\
This work is supported in part by Financiadora de Estudos e Projetos (FINEP) under Project Number 01.10.0233.00, and by Funda\c{c}\~ao de Amparo \`a Pesquisa do Estado de S\~ao Paulo (FAPESP) under Process Number 2012/23767-2.

\normalsize
\nocite{*}
\bibliographystyle{eptcs}
\bibliography{generic}

\begin{thebibliography}{10}
\providecommand{\bibitemdeclare}[2]{}
\providecommand{\surnamestart}{}
\providecommand{\surnameend}{}
\providecommand{\urlprefix}{Available at }
\providecommand{\url}[1]{\texttt{#1}}
\providecommand{\href}[2]{\texttt{#2}}
\providecommand{\urlalt}[2]{\href{#1}{#2}}
\providecommand{\doi}[1]{doi:\urlalt{http://dx.doi.org/#1}{#1}}
\providecommand{\bibinfo}[2]{#2}

\bibitemdeclare{inproceedings}{anderson1996model}
\bibitem{anderson1996model}
\bibinfo{author}{Richard~J \surnamestart Anderson\surnameend},
  \bibinfo{author}{Paul \surnamestart Beame\surnameend}, \bibinfo{author}{Steve
  \surnamestart Burns\surnameend}, \bibinfo{author}{William \surnamestart
  Chan\surnameend}, \bibinfo{author}{Francesmary \surnamestart
  Modugno\surnameend}, \bibinfo{author}{David \surnamestart Notkin\surnameend}
  \& \bibinfo{author}{Jon~D \surnamestart Reese\surnameend}
  (\bibinfo{year}{1996}): \emph{\bibinfo{title}{Model checking large software
  specifications}}.
\newblock In: {\sl \bibinfo{booktitle}{ACM SIGSOFT Software Engineering
  Notes}}, \bibinfo{volume}{21}, \bibinfo{organization}{ACM}, pp.
  \bibinfo{pages}{156--166}, \doi{10.1145/239098.239127}.

\bibitemdeclare{book}{baier/08}
\bibitem{baier/08}
\bibinfo{author}{C.~\surnamestart Baier\surnameend} \& \bibinfo{author}{J.-P.
  \surnamestart Katoen\surnameend} (\bibinfo{year}{2008}):
  \emph{\bibinfo{title}{Principles of model checking}}.
\newblock \bibinfo{publisher}{MIT Press}, \bibinfo{address}{Cambridge, MA,
  USA}.
\newblock
  \urlprefix\url{http://mitpress.mit.edu/books/principles-model-checking}.
\newblock \bibinfo{note}{975 p.}

\bibitemdeclare{inproceedings}{baresi/12}
\bibitem{baresi/12}
\bibinfo{author}{L.~\surnamestart Baresi\surnameend},
  \bibinfo{author}{A.~\surnamestart Morzenti\surnameend},
  \bibinfo{author}{A.~\surnamestart Motta\surnameend} \&
  \bibinfo{author}{M.~\surnamestart Rossi\surnameend} (\bibinfo{year}{2012}):
  \emph{\bibinfo{title}{Towards the UML-Based Formal Verification of Timed
  Systems}}.
\newblock In: {\sl \bibinfo{booktitle}{F. Meth. for Components and Objects}},
  \bibinfo{organization}{Springer}, pp. \bibinfo{pages}{267--286},
  \doi{10.1007/978-3-642-25271-6}.

\bibitemdeclare{article}{behrmann2004tutorial}
\bibitem{behrmann2004tutorial}
\bibinfo{author}{G.~\surnamestart Behrmann\surnameend},
  \bibinfo{author}{A.~\surnamestart David\surnameend} \&
  \bibinfo{author}{K.~\surnamestart Larsen\surnameend} (\bibinfo{year}{2004}):
  \emph{\bibinfo{title}{A tutorial on uppaal}}.
\newblock {\sl \bibinfo{journal}{Formal methods for the design of real-time
  systems}}, pp. \bibinfo{pages}{33--35}, \doi{10.1007/978-3-540-30080-9-7}.

\bibitemdeclare{book}{cockburn}
\bibitem{cockburn}
\bibinfo{author}{Alistair \surnamestart Cockburn\surnameend}
  (\bibinfo{year}{2000}): \emph{\bibinfo{title}{Writing Effective Use Cases}},
  \bibinfo{edition}{1st} edition.
\newblock \bibinfo{publisher}{Addison-Wesley Longman Publishing Co., Inc.},
  \bibinfo{address}{Boston, MA, USA}.
\newblock \urlprefix\url{http://dl.acm.org/citation.cfm?id=517669}.

\bibitemdeclare{article}{cortellessa2002prima}
\bibitem{cortellessa2002prima}
\bibinfo{author}{V.~\surnamestart Cortellessa\surnameend} \&
  \bibinfo{author}{R.~\surnamestart Mirandola\surnameend}
  (\bibinfo{year}{2002}): \emph{\bibinfo{title}{PRIMA-UML: a performance
  validation incremental methodology on early UML diagrams}}.
\newblock {\sl \bibinfo{journal}{SC Programming}}
  \bibinfo{volume}{44}(\bibinfo{number}{1}), pp. \bibinfo{pages}{101--129},
  \doi{10.1016/S0167-6423(02)00033-3}.

\bibitemdeclare{book}{debbabi/10}
\bibitem{debbabi/10}
\bibinfo{author}{M.~\surnamestart Debbabi\surnameend},
  \bibinfo{author}{F.~\surnamestart Hassa{\"i}ne\surnameend},
  \bibinfo{author}{Y.~\surnamestart Jarraya\surnameend},
  \bibinfo{author}{A.~\surnamestart Soeanu\surnameend} \&
  \bibinfo{author}{L.~\surnamestart Alawneh\surnameend} (\bibinfo{year}{2010}):
  \emph{\bibinfo{title}{Verification and Validation in Systems Engineering}}.
\newblock \bibinfo{publisher}{Springer}, \bibinfo{address}{Berlin, Heidelberg -
  Germany}, \doi{10.1007/978-3-642-15228-3}.
\newblock \bibinfo{note}{270 p.}

\bibitemdeclare{inproceedings}{dubrovin2008symbolic}
\bibitem{dubrovin2008symbolic}
\bibinfo{author}{J.~\surnamestart Dubrovin\surnameend} \&
  \bibinfo{author}{T.~\surnamestart Junttila\surnameend}
  (\bibinfo{year}{2008}): \emph{\bibinfo{title}{Symbolic model checking of
  hierarchical UML state machines}}.
\newblock In: {\sl \bibinfo{booktitle}{Application of Concurrency to System
  Design, 2008. 8th International Conference on}},
  \bibinfo{organization}{IEEE}, pp. \bibinfo{pages}{108--117},
  \doi{10.1109/ACSD.2008.4574602}.

\bibitemdeclare{inproceedings}{dwyer/99}
\bibitem{dwyer/99}
\bibinfo{author}{M.~B. \surnamestart Dwyer\surnameend}, \bibinfo{author}{G.~S.
  \surnamestart Avrunin\surnameend} \& \bibinfo{author}{J.~C. \surnamestart
  Corbett\surnameend} (\bibinfo{year}{1999}): \emph{\bibinfo{title}{Patterns in
  property specifications for finite-state verification}}.
\newblock In: {\sl \bibinfo{booktitle}{Proceedings...}},
  \bibinfo{organization}{International Conference on Software Engineering
  (ICSE)}, \bibinfo{publisher}{ACM}, \bibinfo{address}{New York, NY, USA}, pp.
  \bibinfo{pages}{411--420}, \doi{10.1145/302405.302672}.

\bibitemdeclare{article}{eshuis2006symbolic}
\bibitem{eshuis2006symbolic}
\bibinfo{author}{R.~\surnamestart Eshuis\surnameend} (\bibinfo{year}{2006}):
  \emph{\bibinfo{title}{Symbolic model checking of UML activity diagrams}}.
\newblock {\sl \bibinfo{journal}{ACM Transactions on Software Engineering and
  Methodology (TOSEM)}} \bibinfo{volume}{15}(\bibinfo{number}{1}), pp.
  \bibinfo{pages}{1--38}, \doi{10.1145/1125808.1125809}.

\bibitemdeclare{book}{holzmann2004spin}
\bibitem{holzmann2004spin}
\bibinfo{author}{G.J. \surnamestart Holzmann\surnameend}
  (\bibinfo{year}{2004}): \emph{\bibinfo{title}{The SPIN model checker: Primer
  and reference manual}}.
\newblock \bibinfo{volume}{1003}, \bibinfo{publisher}{Addison-Wesley}.
\newblock \urlprefix\url{http://spinroot.com/spin/Doc/Book_extras/}.

\bibitemdeclare{inproceedings}{johnson2013incremental}
\bibitem{johnson2013incremental}
\bibinfo{author}{K.~\surnamestart Johnson\surnameend},
  \bibinfo{author}{R.~\surnamestart Calinescu\surnameend} \&
  \bibinfo{author}{S.~\surnamestart Kikuchi\surnameend} (\bibinfo{year}{2013}):
  \emph{\bibinfo{title}{An Incremental Verification Framework for
  Component-based Software Systems}}.
\newblock In: {\sl \bibinfo{booktitle}{Proceedings of the 16th International
  ACM Sigsoft Symposium on Component-based Software Engineering}},
  \bibinfo{series}{CBSE '13}, \bibinfo{publisher}{ACM}, \bibinfo{address}{New
  York, NY, USA}, pp. \bibinfo{pages}{33--42}, \doi{10.1145/2465449.2465456}.

\bibitemdeclare{misc}{NuSMV}
\bibitem{NuSMV}
\bibinfo{author}{FONDAZIONE~BRUNO \surnamestart KESSLER\surnameend}
  (\bibinfo{year}{2011}): \emph{\bibinfo{title}{NuSMV Home Page}}.
\newblock \urlprefix\url{http://nusmv.fbk.eu/}.

\bibitemdeclare{article}{lam2007formalism}
\bibitem{lam2007formalism}
\bibinfo{author}{Vitus S.~W. \surnamestart Lam\surnameend}
  (\bibinfo{year}{2007}): \emph{\bibinfo{title}{A Formalism for Reasoning About
  UML Activity Diagrams}}.
\newblock {\sl \bibinfo{journal}{Nordic J. of Computing}}
  \bibinfo{volume}{14}(\bibinfo{number}{1}), pp. \bibinfo{pages}{43--64}.
\newblock \urlprefix\url{http://dl.acm.org/citation.cfm?id=1515784.1515786}.

\bibitemdeclare{inproceedings}{merseguer2002compositional}
\bibitem{merseguer2002compositional}
\bibinfo{author}{J.~\surnamestart Merseguer\surnameend},
  \bibinfo{author}{J.~\surnamestart Campos\surnameend},
  \bibinfo{author}{S.~\surnamestart Bernardi\surnameend} \&
  \bibinfo{author}{S.~\surnamestart Donatelli\surnameend}
  (\bibinfo{year}{2002}): \emph{\bibinfo{title}{A Compositional Semantics for
  UML State Machines Aimed at Performance Evaluation}}.
\newblock \bibinfo{series}{WODES '02}, \bibinfo{publisher}{IEEE Computer
  Society}, \bibinfo{address}{Washington, DC, USA}, pp. \bibinfo{pages}{295--}.
\newblock \urlprefix\url{http://dl.acm.org/citation.cfm?id=832316.837588}.

\bibitemdeclare{inproceedings}{mikk1998implementing}
\bibitem{mikk1998implementing}
\bibinfo{author}{E.~\surnamestart Mikk\surnameend},
  \bibinfo{author}{Y.~\surnamestart Lakhnech\surnameend},
  \bibinfo{author}{M.~\surnamestart Siegel\surnameend} \& \bibinfo{author}{G.J.
  \surnamestart Holzmann\surnameend} (\bibinfo{year}{1998}):
  \emph{\bibinfo{title}{Implementing statecharts in PROMELA/SPIN}}.
\newblock In: {\sl \bibinfo{booktitle}{Industrial Strength Formal Specification
  Techniques, 1998.}}, \bibinfo{organization}{IEEE}, pp.
  \bibinfo{pages}{90--101}, \doi{10.1109/WIFT.1998.766303}.

\bibitemdeclare{misc}{umlref}
\bibitem{umlref}
\bibinfo{author}{The Object Management~Group \surnamestart OMG\surnameend}
  (\bibinfo{year}{1997}): \emph{\bibinfo{title}{OMG - Unified Modeling Language
  (OMG UML)}}.
\newblock \urlprefix\url{http://www.uml.org/}.

\bibitemdeclare{article}{santiago/12}
\bibitem{santiago/12}
\bibinfo{author}{V.~A. \surnamestart {Santiago J\'unior}\surnameend} \&
  \bibinfo{author}{N.~L. \surnamestart {Vijaykumar}\surnameend}
  (\bibinfo{year}{2012}): \emph{\bibinfo{title}{Generating Model-Based Test
  Cases from Natural Language Requirements for Space Application Software}}.
\newblock {\sl \bibinfo{journal}{Software Quality Journal}}
  \bibinfo{volume}{20}(\bibinfo{number}{1}), pp. \bibinfo{pages}{77--143},
  \doi{10.1007/s11219-011-9155-6}.

\bibitemdeclare{article}{schafer2001model}
\bibitem{schafer2001model}
\bibinfo{author}{T.~\surnamestart Sch{\"a}fer\surnameend},
  \bibinfo{author}{A.~\surnamestart Knapp\surnameend} \&
  \bibinfo{author}{S.~\surnamestart Merz\surnameend} (\bibinfo{year}{2001}):
  \emph{\bibinfo{title}{Model checking UML state machines and collaborations}}.
\newblock {\sl \bibinfo{journal}{Electronic Notes in Theoretical Computer
  Science}} \bibinfo{volume}{55}(\bibinfo{number}{3}), pp.
  \bibinfo{pages}{357--369}, \doi{10.1016/S1571-0661(04)00262-2}.

\bibitemdeclare{inproceedings}{uchitel2001workbench}
\bibitem{uchitel2001workbench}
\bibinfo{author}{Sebastian \surnamestart Uchitel\surnameend} \&
  \bibinfo{author}{Jeff \surnamestart Kramer\surnameend}
  (\bibinfo{year}{2001}): \emph{\bibinfo{title}{A workbench for synthesising
  behaviour models from scenarios}}.
\newblock In: {\sl \bibinfo{booktitle}{Proceedings of the 23rd intern.
  conference on Software engineering}}, \bibinfo{organization}{IEEE Computer
  Society}, pp. \bibinfo{pages}{188--197}, \doi{10.1109/ICSE.2001.919093}.

\bibitemdeclare{article}{woodcock/09}
\bibitem{woodcock/09}
\bibinfo{author}{J.~\surnamestart Woodcock\surnameend}, \bibinfo{author}{P.~G.
  \surnamestart Larsen\surnameend}, \bibinfo{author}{J.~\surnamestart
  Bicarregui\surnameend} \& \bibinfo{author}{J.~\surnamestart
  Fitzgerald\surnameend} (\bibinfo{year}{2009}): \emph{\bibinfo{title}{Formal
  methods: Practice and experience}}.
\newblock {\sl \bibinfo{journal}{ACM Computing Surveys}}
  \bibinfo{volume}{41}(\bibinfo{number}{4}), pp. \bibinfo{pages}{19:1--19:36},
  \doi{10.1145/1592434.1592436}.

\end{thebibliography}

\end{document}